\documentstyle[preprint,aps]{revtex}
\def\CQG{ Class. Quant. Grav. }

\def\PLB{ Phys. Lett.  }

\def\JHEP{ J. High Energy Phys. }
\begin{document}
\preprint{\vbox{\noindent
\null\hfill INFNCA-TH0006}}
%
%
\draft
\title{A Realization of the infinite-dimensional Symmetries of
Conformal Mechanics}
\author{Mariano Cadoni$^{a,c,}$\footnote{E--Mail: cadoni@ca.infn.it},
Paolo Carta$^{a,c,}$\footnote{E--Mail: carta@ca.infn.it} and Salvatore
Mignemi$^{b,c,}$\footnote{E--Mail: mignemi@ca.infn.it}}
\address{$^a$ Dipartimento di Fisica, Universit\`a di Cagliari,\\
Cittadella Universitaria 09042, Monserrato, Italy.}
\address{$^b$ Dipartimento di Matematica, Universit\`a di Cagliari,\\
viale Merello 92, 09123, Cagliari, Italy.}
\address{$^c$ INFN, Sezione di Cagliari}
\maketitle
\begin{abstract}
We discuss the possibility of realizing the infinite dimensional
symmetries of conformal mechanics as time reparametrizations,
generalizing the realization of the $SL(2,\mathbf{R})$ symmetry of the
de Alfaro, Fubini, Furlan model in terms of quasi--primary fields.  We
find that this is possible using an appropriate generalization of the
transformation law for the quasi--primary fields.
\end{abstract}
\section{Introduction}
Recently, much attention has been devoted to the study of the
symmetries of conformal mechanics \cite{kumar,caccia,cm}, especially
because this model is expected to play an important role in the
context of the $\rm AdS_2/CFT_1$ correspondence
\cite{cm,isoliti}. Another reason of the renewed interest in this
model is the fact that it describes the motion of a particle near the
horizon of the Reissner-Nordstrom black hole \cite{kallosh}.

Starting from the symplectic symmetry of its phase space, it was shown
that conformal mechanics admits an infinite set of conserved charges,
which span a $w_\infty$ algebra \cite{kumar,caccia}. However, the
physical interpretation of this algebra and of the related charges,
and its connection with the conformal algebra of the original de
Alfaro, Fubini, Furlan (DFF) model \cite{dff} has not been discussed
in the literature.  In particular, the Virasoro algebra appearing as
subgroup of the $w_\infty$ algebra can be thought of as an infinite
dimensional extension of the $SL(2,\mathbf{R})$ symmetry of the DFF
model. It seems, therefore, very natural that a realization of the
infinite dimensional symmetry of conformal mechanics exists, which
generalizes the realization of the $SL(2,\mathbf{R})$ symmetry of the
DFF model in terms of quasi--primary fields.  The $SL(2,\mathbf{R})$
symmetry of the DFF model is realized as conformal transformations of
the time $t$ with the coordinate $q$ transforming as a quasi-primary
conformal field \cite{dff}.  The most natural generalization of this
realization leading to an infinite dimensional symmetry is obtained by
requiring invariance under diffeomorphisms of the time $t$.  Such a
generalization is also consistent with the fact that the conformal
group in one dimension can be realized as the group of time
reparametrizations and is generated by a Virasoro algebra \cite{cm}.
An implementation of the previous ideas could help to shed some light
on the mysteries of the AdS$_{2}$/CFT$_{1}$ correspondence.  

In this paper we discuss in detail these points. In particular, we
show how the realization of the $SL(2,\mathbf{R})$ symmetry of the DFF
model can be generalized to an infinite dimensional symmetry realized
as time reparametrizations.  The price we pay in doing this operation
is that the transformation law for the coordinate $q$ is not given by
the simple transformation law of a conformal field with given weight
but contains further terms depending on the dynamical variables of the
system.

Another, related, issue we will address in this paper is to understand
how a infinite dimensional symmetry can arise in a model with only one
degree of freedom.  We will show that the $w_\infty$ and conformal
algebras have different origin, and that the generators of $w_\infty$
are not functionally independent, but are functions of two basic
conserved quantities.

The structure of the paper is the following. In sect. 2 we discuss the
conformal symmetry of conformal mechanics and its realization in terms
of time reparametrizations generalizing the DFF realization of the
$SL(2,\mathbf{R})$ symmetry . In sect. 3, we show how this symmetry
can be further extended to the full $w_\infty$ algebra.

\section{Symmetries of one--dimensional Conformal Mechanics}
In Ref. \cite{kumar} was observed that the algebra of the symmetries
of a generic one--dimensional conformal model described by the
Hamiltonian,
\begin{equation}\label{e:ham}
H=\frac{p^2}{2f(u)},
\end{equation}
where $f(u)$ is a function of $u=pq$,  can be extended to a Virasoro
algebra with generators
\begin{displaymath}
L_n=-\frac{1}{2}q^{1+n}p^{1-n}f^n,
\end{displaymath}
such that\footnote{With $\{F,G\}$ we mean the Poisson Brackets with
the convention $\{q,p\}=1$.}

\begin{displaymath}
\{L_n,L_m\} = (m-n)L_{m+n}.
\end{displaymath}
The generators $L_n$ are not conserved charges.  The origin of the
symmetry is rather obscure in this formulation, but was partially
clarified in Ref. \cite{caccia}, where the symmetry was also extended
to a $w_\infty$ algebra. Since the Hamiltonian (\ref{e:ham}) can be
transformed into the free Hamiltonian $H=p^2/2$ by a simple canonical
transformation, the symmetries of the action are easily found. It
turns out that the diffeomorphisms which preserve the symplectic form
$\Omega = dp \wedge dq -dH \wedge dt$ are generated by vector fields
spanning the $w_\infty$ algebra of area--preserving
diffeomorphisms\cite{sez}. One can choose a basis of generators
\begin{equation}\label{e:gen}
v^l_m = \left( \frac{p}{\sqrt{f}}\right)^{l+1} \left( q\sqrt{f}
-\frac{p}{\sqrt{f}}t\right)^{l+m+1},
\end{equation}
with $l=0,1,2,\ldots$ and $m\in\mathbf{Z}$, which
satisfy the $w_\infty$ algebra:
\begin{equation}\label{e:winf}
\{ v_m^l, v_{m'}^{l'}\} = [m(l'+1)-m'(l+1)]v_{m+m'}^{l+l'}.
\end{equation}
The $v_m^l$ are conserved charges (constants of motion):
\begin{equation}
\frac{d v_m^l}{dt} = \frac{\partial v_m^l}{\partial t} + \{ v_m^l, H\}
=0.
\end{equation}

The physical interpretation of these charges is still not completely
evident. However, the same result can be recovered in a more intuitive
way, when one tries to extend to an infinite dimensional Virasoro
algebra the $SL(2,\mathbf{R})$ symmetry algebra of
the conformal mechanics of DFF \cite{dff},  whose  Hamiltonian
\begin{equation}\label{e:dff}
H_{DFF} = \frac{p^2}{2} +\frac{g}{2q^2},
\end{equation}
with $g$ constant, is a special case of the model (1) with
$f=(1+g/u^2)^{-1}$.  The coordinate $q$ transforms under the conformal
group $SL(2,\mathbf{R})$ as
a quasi--primary field:
\begin{equation}\label{e:primary}
q'(t') = \frac{1}{\gamma t + \delta}\, q(t), \qquad t'=\frac{\alpha t
+\beta}{\gamma t + \delta},\qquad \alpha\delta-\beta\gamma =1.
\end{equation}
The generators of infinitesimal translations, dilatations and special
conformal transformations are given by $H_{DFF}$, $D_{DFF}$ and
$K_{DFF}$, where
\begin{eqnarray}\label{e:cariche_dff}
&& D_{DFF} = tH -\frac{pq}{2}\ ,\nonumber\\
&& K_{DFF} = t^2H -tpq + \frac{q^2}{2}\ .
\end{eqnarray}
These generators are conserved under time evolution.
The conformal $SL(2,\mathbf{R})$ algebra is
\begin{equation}\label{e:a_picc}
\{H_{DFF}, D_{DFF} \}=H_{DFF}, \quad \{K_{DFF},D_{DFF} \}=-K_{DFF},
\quad \{H_{DFF},K_{DFF} \}=2D_{DFF}.
\end{equation}

One can show that it is impossible to find an infinite dimensional
extension of the little algebra generated by $H_{DFF}=L_{-1}$,
$D_{DFF}=L_0$, $K_{DFF}=L_1$, which is realized as time
reparametrizations and preserves the transformation law for the
coordinates $q$ given in Eq. (\ref{e:primary}).  In fact, since $\{
L_{-1},L_0,L_1\}$ are conserved charges, any extension can consist
only of conserved charges. This proposition is easily proven. From
\begin{eqnarray}\label{e:dim}
\{L_n,L_{-1}\}&&=-(n+1)L_{n-1}\nonumber\\
\{L_n,L_{1}\}&&=-(n-1)L_{n+1}\nonumber\\
\frac{dL_{-1}}{dt}=\frac{dL_{0}}{dt}&&=\frac{dL_{1}}{dt}=0,
\end{eqnarray}
we conclude that
\begin{equation}
\frac{dL_{-2}}{dt}=\frac{dL_{2}}{dt}=0 \quad \Rightarrow \quad
\frac{dL_{n}}{dt}=0, \quad n\le -2 \quad \mbox{and} \quad n\ge +2.
\end{equation}
The first relation on the r.h.s. of the previous implication is proved
considering the total time derivative of
\begin{displaymath}
\{L_{2},L_{-1}\}=-3L_{1}
\end{displaymath}
and the last line of (\ref{e:dim}). From $\frac{d}{dt}L_n =0$ for $n
\ge -1$, we infer that also the remaining generators $L_n$, $n \le -2$
are conserved. But this conclusion implies that the coordinate $q$
should transform as a primary (not only quasi--primary) field in order
to give a conserved action modulo a total time derivative, which is
not the case.  In fact, if we consider a generic time
reparametrization $t' = h(t)$ we have
\begin{equation}\label{diff}
q'(t') = \left(\frac{dh(t)}{dt}\right)^{1/2} q(t)\ .
\end{equation}
The variation of the action $S = \int dt L(q,\dot{q})$ is given by
\begin{equation}\label{e:ds}
\delta S = \frac{1}{2}\int dt\, \left[ \frac{d}{dt}\left( \frac{\ddot
h}{\dot h}\,\frac{q}{2}\right)- (h,t)\frac{q^2}{2}\right],
\end{equation}
where $(h,t)$ is the Schwarzian derivative of the map $t\to
t'=h(t)$. It is easy to see that $\delta S$ is a total derivative only
if $h(t)$ is a conformal map, while for a generic infinitesimal time
transformation $t' = t+\epsilon g(t)$, Eq. (\ref{e:ds}) becomes:
\begin{equation}
\delta S = \frac{1}{2}\int dt\, \epsilon \ddot g q\dot q\ .
\end{equation}
The conclusion is that there cannot exist conserved charges
corresponding to time-reparametrizations with the field $q$
transforming as in Eq.  (\ref{diff}).  This implies, in turn, that it
is impossible to generalize the $SL(2,\mathbf{R})$ conformal symmetry
(\ref{diff}) to the full diffeomorphisms group, if one requires that
$q$ transforms as a conformal field.

It is evident that the quasi--primarity condition (\ref{e:primary}) is
too strong. Let us try to relax this condition and impose an analogous
condition only for translations and dilatations.  From Noether theorem
we get $D=tH -pq/2$ $(=D_{DFF})$, which in turn implies
\begin{equation}
\frac{d D}{dt} = 0 \Rightarrow H = \frac{p^2}{2f(pq)}.
\end{equation}

Hence, $H$ must have the form given by Eq. (\ref{e:ham}) with $f(pq)$
being an arbitrary
function. If we further impose that $q$ transforms as in
(\ref{e:primary}) (quasi--primary) for infinitesimal special conformal
transformation of $t$, we are forced to take
\begin{eqnarray}\label{e:forced}
H && = \frac{p^2}{2}\left( 1 + \frac{g_1}{pq} +
\frac{g_2}{(pq)^2}\right)\nonumber \\
K && = t^2H -tpq + \frac{q^2}{2} -\frac{g_1}{2}t,
\end{eqnarray}
where $g_1$ and $g_2$ are constants.  With the trivial canonical
transformation $p' = p+g_1/(2q)$, $q' = q$, we have $H \to H_{DFF}$
with $g=g_2-g_1^2/4$, so the quasi--primarity condition for the $q$ is
fulfilled only by the DFF conformal mechanics. Let us assume instead
that under infinitesimal special conformal transformations $t' = t
-\omega t^2$, $q$ changes according to
\begin{equation}
\delta_0 q = \omega\{q,K\} =
\omega t^2 \dot q -\omega t q + \,\,\text{corrections}
\end{equation}
where $\delta_0 q = q'(t)-q(t)$ and $K$ is the (conserved) charge to
be determined. If we take $K = t^2H -tpq +f_2$, the condition
$\frac{d}{dt}K =0$ translates into
\begin{equation}
\frac{d}{dt}f_2 = pq  \Rightarrow \left\{ \begin{array}{ll}
f_2 = \frac{q^2}{2} & \mbox{DFF solution}\\
f_2 = \frac{(pq)^2}{4H} &
\end{array}
\right.
\end{equation}

If we choose the second solution, we can write the three conserved
generators as:
\begin{eqnarray}\label{e:nostri}
L_{-1} && = H_{CCM} = H\left(t-\frac{pq}{2H}\right)^0, \nonumber \\
L_{0} && = D_{CCM} = H\left(t-\frac{pq}{2H}\right)^1 ,\nonumber \\
L_{1} && = K_{CCM} = H\left(t-\frac{pq}{2H}\right)^2.
\end{eqnarray}

The extension to a Virasoro Algebra of conserved charges is now
immediate:
\begin{eqnarray}\label{gener}
&& L_{n} =  H\left(t-\frac{pq}{2H}\right)^{1+n} \qquad n \in
\mathbf{Z} \nonumber \\
&& \frac{d}{dt} L_{n} = 0 \nonumber \\
&& \{L_{n},L_{m}\} = (m-n)L_{n+m}.
\end{eqnarray}

The meaning of these charges is clear from their very construction:
they canonically implement the transformation of the variables $q,p$
as primary ``fields'' plus interaction-dependent corrections necessary
to close a Virasoro algebra of conserved quantities. It can be noted
that the DFF generators of the $SL(2,\mathbf{R})$ algebra coincide with
the generators (\ref{e:nostri}) only in the non-interacting $g=0$
case.

According to this interpretation we can establish a correspondence
between the charges $L_{n-1}$ and the infinitesimal time
transformations $t \to t -\omega t^n$, with $n\in\mathbf{Z}$,
identifying the $L_{n-1}$ with the Noether charges for such
transformations.  In fact, under $L_{n-1}$ the coordinates and momenta
transform as

\begin{eqnarray}\label{tra}
\delta_{0} q &&= \omega \{q, L_{n-1}\}\nonumber \\
\delta_{0} p &&= \omega \{p, L_{n-1}\}.
\end{eqnarray}
Using Eq. (\ref{tra}) one can easily find the transformation law of
$p,q$ for a generic infinitesimal time transformation $t' = t
-\epsilon(t)$,
\begin{eqnarray}
\delta_0 q &&=  \{q, H\epsilon\left( t
-\frac{pq}{2H}\right)\}\nonumber\\
\delta_0 p &&=  \{p, H\epsilon\left( t -\frac{pq}{2H}\right)\}.
\end{eqnarray}
Expanding $\epsilon(t)=\sum_n\epsilon_n t^n$, and computing the Poisson
brackets, we get
\begin{eqnarray}
\delta_0 q &&= \left(\epsilon \dot q-{1\over 2} \dot\epsilon q\right)+
\sum_n\epsilon_n\sum_{k\ge 2}{n\choose k} t^{n-k}
\left\{q,H\left(-\frac{pq}{2H}\right)^{k}\right\},\nonumber \\
\delta_0 p &&=
\left(\epsilon \dot p+{1\over 2} \dot\epsilon p\right)+
\sum_n\epsilon_n\sum_{k\ge 2}{n\choose k} t^{n-k}
\left\{p,H\left(-\frac{pq}{2H}\right)^{k}\right\}.
\end{eqnarray}
Comparing this realization of the infinite dimensional conformal
symmetry with the usual realization in terms of primary fields,
\begin{equation}
\delta_0 q = \epsilon \dot q-{1\over 2} \dot\epsilon q\ ,
\end{equation}
one easily sees that our realization of the infinite dimensional
conformal algebra differs from the usual one for terms depending on
the dynamics of the system.  In particular, in contrast with ordinary
"spacetime" symmetries, the additional terms depend both on $q$ and
$\dot q$.

\section{Further extensions of the symmetry algebra}
The Virasoro algebra (\ref{gener}) can be easily extended to a
$w_\infty$-algebra analogous to (\ref{e:winf}). In fact, starting from
the two elementary conserved charges $H$ and $G= t-\frac{pq}{2H}$, one
can define the conserved generators
\begin{displaymath}
w_{l,m} = H^{1+l}G^m \qquad l,m\in\mathbf{Z},
\end{displaymath}
with Poisson brackets
\begin{equation}\label{mycharges}
\{w_{l,m},w_{l',m'}\} = [m'(l+1)-m(l'+1)]
\,w_{l+l',m+m'}.
\end{equation}
The charges so defined satisfy the same algebra as (\ref{e:winf}),
and essentially coincide with the charges $v^l_m$, modulo some
redefinitions. The Virasoro algebra
(\ref{gener}) is contained in (\ref{mycharges}) as a
subalgebra, $L_n=w_{0,n+1}$.

It should be noticed that in the particular case of DFF conformal
mechanics, the algebra (\ref{mycharges}) also includes the generator
of special conformal transformations $K_{DFF}$. In fact,
from the definitions (\ref{e:cariche_dff}) and (\ref{gener}),
\begin{equation}
K_{DFF} = K_{CCM} + \frac{g}{4H_{DFF}} = w_{0,2}+{g\over4}w_{-2,0}.
\end{equation}

Our construction clarifies how an infinite set of conserved charges
can originate from a system with only one degree of
freedom: all the charges are in fact functionally dependent from the
two elementary charges $H$ and $G$.

\end{document}